\documentstyle[12pt]{article}
\begin{document}

\begin{titlepage}

\hfill {DOE/ER/40322-154}
\hfill {U. of MD PP \#92-191}
\vspace{20pt}

\centerline{\bf The role of the Delta isobar in chiral perturbation theory}
\centerline{\bf and hedgehog soliton models}

\vspace{20pt}

\centerline{Thomas D.  Cohen and Wojciech Broniowski
\footnote{\noindent On leave of absence from H. Niewodnicza\'nski
Institute of Nuclear Physics, ul. Radzikowskiego 152, 31-342 Cracow, POLAND} }

\centerline{\it Department of Physics and Astronomy}
\centerline{{\it University of Maryland, College Park, Maryland}
{\sl 20742-4111}}

\vspace{32pt}

\centerline{Abstract}
{\noindent

Hedgehog model predictions for the leading
nonanalytic behavior (in $m^{2}_{\pi }$) of certain observables are
shown to agree with the predictions of chiral perturbation theory up to
an overall factor which
depends on the operator. This factor can be understood in terms of
contributions of the $\Delta$ isobar in chiral loops.  These
physically motivated  contributions are
analyzed in an expansion in which both $m_{\pi}$ and $M_{\Delta}-M_N$
are taken as small parameters, and are shown to
yield large corrections to both  hedgehog models and chiral perturbation
theory.}

\end{titlepage}

There is a wide spread consensus that the pion cloud plays an important
role in the structure of the nucleon --- the pion is very light and therefore
very long ranged.  Moreover, the pion is rather well understood in terms of
spontaneous chiral symmetry breaking of the underlying theory, QCD.
Accordingly, the pion is emphasized in a large number of
hedgehog soliton models of the
nucleon, including various incarnations of the Skyrme model
\cite{skyrmion:rev}, the chiral
or hybrid bag model \cite{hybrid:rev}, the chiral quark-meson
model \cite{BBC:rev}, the chiral
color-dielectric model \cite{conf}, or the Nambu--Jona-Lasinio model
in the solitonic treatment \cite{NJL:sol}.

Apart from these models, there is another approach to the structure of
hadrons which emphasizes the role of pions, namely
chiral perturbation theory $(\chi PT)$.
Its basic premise is that there is a separation of scales between the pion
mass and all other mass scales in the problem (this separation becomes
increasingly good as one approaches the chiral limit).  Low momentum
observables are studied via systematic expansion in $m_{\pi}^2$
(or equivalently the quark mass).

In this note we study the
relationship between $\chi PT$ and  hedgehog soliton models
Our central point is that for a certain class of
observables (those whose long range physics is dominated by pion bilinears and
which which are scalar-isoscalar or vector-isovector) the hedgehog
models agree with the predictions of $(\chi PT)$ for the leading
nonanalytic behavior (in $m^{2}_{\pi })$ {\it up to an overall factor
which depends on the quantum numbers of the operator}.
This factor, equal to 3 for scalar-isoscalar operators, and 3/2
for vector-isovector operators,  can be traced to the
noncommutativity of the large-$N_c$ (number of colors) limit and the chiral
limit. The essential physics behind this noncommutativity
is the role of the $\Delta$ resonance.
The $\Delta$ makes important contributions in the hedgehog models, where it
is treated as degenerate with the nucleon. On the other hand, its
contributions to the leading
nonanalytic behavior are not included in conventional $\chi PT$, since the
$N$-$\Delta$ splitting is assumed to be much larger than $m_{\pi}$.
We analyze the role of the $\Delta$ in chiral loops in the
spirit of Ref. \cite{JM}, taking
physical values for the $N$-$\Delta$ splitting and $m_{\pi}$, and find
large corrections to both  hedgehog models and chiral perturbation theory.
This suggests how hedgehog results and $\chi PT$ results should be
corrected to account for effects finite $N$-$\Delta$ splitting.

The hedgehog soliton models are designed
to be used at the mean-field level, which can be justified in
the large-$N_c$ limit of QCD \cite{witten:nc}.  Stable
mean-field configurations are hedgehog solitons
in which the internal isospin index is correlated with the spatial index.  For
pions, the hedgehog configuration is $\pi_a = f(r) \hat{r}_a$, where $a$ is the
isospin index, $f$ is a spherically symmetric profile function and
$\hat{r}_a$ is a spatial unit vector pointing out from the center of the
soliton.
Appropriate forms can be written for other fields.
As is well known, these hedgehogs break both the rotational and isorotational
symmetries of the model lagrangian, while preserving the ``grand rotational
symmetry'' generated by ${\bf K}={\bf I}+{\bf J}$. As a result, the
hedgehog {\it does not} have quantum numbers of physical baryons.
Instead, it represents a deformed
intrinsic state which corresponds to a {\it band of states}.  Information about
physical states is obtained with a semiclassical projection method
\cite{ANW83,CB86}, which
gives the matrix elements of arbitrary operators in a form which is
manifestly correct {\it to leading order in the} $1/N_c$ {\it expansion}.

In $\chi PT$, beyond the lowest order, one typically  has to include both tree
diagrams and loops (which are suitably cut off at the separation scale)
\cite{Weiberg,GasserL}.
Infrared divergences in the pion loops (at $m^{2}_{\pi }=0$) lead to effects
which are {\it nonanalytic} in $m^{2}_{\pi }$.
Recently the
systematic treatment of $\chi PT$ has been extended to the nucleon sector
\cite{GasserSS}. It is believed that the {\it leading nonanalytic behavior} is
given by a single pion-nucleon loop calculation.

The hedgehog models to leading order in $1/N_c$ are essentially classical so
it is by no means obvious that the physics of quantum pion loops should be
present. Somewhat surprisingly, this is precisely what happens
for a class of observables which have nonanalytic behavior, in particular
for observables which diverge as $m^{-1}_{\pi }$, as will be demonstrated
in examples below. Generally, the connection can be seen as follows:
We start from an expression for the pion-nucleon loop, and perform the
integration over the time-component of the momentum flowing around the loop.
The leading divergence picks up contributions from poles in the pion
propagator(s) only --- the nucleon can be treated
non-relativistically, and its
recoil enters at a subleading level. The resulting expression, after
rewriting it in a Fourier transformed manner,
involves a single spatial integral
of a quadratic expression in Hankel functions (or derivatives thereof).
Explicitly, the pion tail in a soliton has the form
\mbox{$\phi^{asymp.}_a = (3 g_A) /(8 \pi F_{\pi}) c_{ai} \widehat{x}_i
(m_{\pi} + 1/r) exp(- m_{\pi} r) /r$}, and involves the same Hankel function
(the collective variables $c_{ai} = Tr[\tau_a B \tau_i B^{\dagger}]$
are discussed in Refs. \cite{ANW83,CB86}). The
above outline shows there is nothing
mysterious about hedgehog models reproducing some
of the physics of the chiral loops. The
noncommutativity of the large-$N_c$ and chiral limits leads, however, to
a mismatch by a constant, which is the key point discussed below
\footnote{For observables not considered in this paper
  (vector-isoscalar or scalar
  isovector) hedgehog models do not predict correct chiral singularities (e.g.,
  for the electric isovector mean squared radius hedgehogs give $m^{-1}_{\pi}$
  rather than $log(m_{\pi})$). Evaluation of these observables explicitly
  involves the dynamics of cranking (results depend on the moment of inertia).
  In the semiclassical projection one finds rotating solutions to the
  time-dependent classical equations of motion {\it to leading order in}
  $1/N_c$ ({\it slow} rotations); centrifugal stretching and other order
  $1/N_c$ effects are ignored. However, centrifugal effects increase with
  distance from the center of the soliton. The longest distance part of the
  configuration is the region for which the $1/N_c$ approximation does the
  worst job in cranking, {\it i.e.} the difference between the leading order
  $1/N_c$ solution of the classical equations of motion and the exact solution
  increases with distance. In this case the issue of nonocommutativity of the
  large-$N_c$ and chiral limits is much more complicated than for observables
  considered in this paper.}.

Perhaps the most straightforward way to see this is to study some explicit
cases.  Here, we will compare the leading singular behavior as $m_{\pi }
\rightarrow 0$ of
several quantities calculated using standard leading order in $N_c$ hedgehog
model techniques with the same quantities calculated in $\chi PT$ at
one loop.  In the case of the hedgehog models the chirally singular
behavior comes from the
long-range tail of the pion field, $\phi^{asymp.}_a$, whose amplitude
depends on $g_{\pi NN}$ (or $g_{A}$).  We consider
the isovector
mean squared magnetic radius (which is expressed in terms of the matrix element
of a
vector-isovector operator),  and two scalar-isoscalar quantities:
$d^{2} M_N / d(m^{2}_{\pi })^{2} = d (\sigma_{\pi N} / m^{2}_{\pi })/
d(m^{2}_{\pi }) $, and  the isoscalar electric
polarizability \mbox{$\alpha_N =(\alpha_p + \alpha_n)/2$}. All these
quantities  diverge as $1/m_{\pi }$ near the chiral limit.
The hedgehog model expression for $d^{2} M_N / d(m^{2}_{\pi })^{2}$ can
be obtained easily from
the identity \mbox{$d M_N / d(m^{2}_{\pi }) = \mbox{$1\over 2$} <N\mid
\int d^{3}x (\phi^{asymp.})^{2}\mid N>$}, the form for
$<r^{2}>_m^{I=1}$ is given in Refs. \cite{ANW83,CB86},
while the electric polarizability is given in Ref. \cite{pol:ours}.
The $\chi PT$
one-loop expressions for the same expressions can
be extracted from Refs. \cite{BegZapeda,GasserSS,BKM91}.

The hedgehog model expressions and the $\chi PT$ predictions are
compared in Table~I.
We see, as advertised, that for the scalar-isoscalar
quantities the hedgehog model results are a factor of three larger than the
one loop $\chi PT$ predictions while for the vector-isovector quantity
the factor is ${3\over 2}$. The reason for this is
associated with the noncommutativity of the large-$N_c$ and chiral
limits.  In chiral perturbation it is
assumed that the pion is very light compared to all other scales in the
problem, and, consequently, dominant contributions in chiral loops come
from $N$-$\pi$
states, which are the lightest excited states with the appropriate quantum
numbers.  These states become increasingly dominant as the chiral limit is
approached (at least for process which are infrared divergent) and they lead
to nonanalytic behavior in $m^{2}_{\pi }$.

In the large-$N_c$  limit the nucleon is essentially degenerate with
the $\Delta $ isobar (the
mass splitting goes as $1/N_c$). Consequently, the pion is {\it not} much
lighter
than all other scales in the problem and the nucleon-pion states are no
longer the {\it only} light intermediate states in the problem ---
$\Delta$-$\pi$ states
are also light.  In one-pion-loop calculations these $\Delta$-$\pi$
states should also be included.
We now see explicitly the issue of ordering of the limits.  In standard
treatments of hedgehog
models one implicitly takes the large-$N_c$ limit before going to the chiral
limit.  Thus $\Delta $-$\pi$ states are degenerate with the $N$-$\pi$
states and hence are not suppressed due to the mass difference.  In contrast,
conventional $\chi PT$ corresponds to the opposite ordering of limits
(first chiral,
than large-$N_c$), in which a finite energy denominator prevents the
$\Delta$-$\pi$
states from giving rise to chiral singularities.  The relative size of
$\Delta $-$\pi$ contributions in one loop calculations is determined by the
relative
strength of the $\pi$-$N$-$\Delta$ coupling to the
the $\pi$-$N$-$N$ coupling,  and by the $N$-$\Delta$ mass difference.
In hedgehog models at large-$N_c$, one has $M_N = M_{\Delta }$,
and $g_{\pi N\Delta } =3/2 \; g_{\pi NN}$ (with the normalization of
Ref. \cite{ANW83}). Then, it is
straightforward to see that the contribution of
the $\Delta$-$\pi$ loop to a nucleon
matrix element of some operator ${\cal O}$ is
just a numerical coefficient times the nucleon contribution.  This
coefficient is determined simply from Clebsch-Gordan algebra,  and depends on
the quantum numbers of ${\cal O}$.  For our vector-isovector and
scalar-isoscalar cases we find
\begin{eqnarray}
\frac{<N| {\cal O}^{I=J}|N>^{1-loop}_{\Delta}}
     {<N| {\cal O}^{I=J}|N>^{1-loop}_{N}} &=& C_{\cal O} , \nonumber \\
 C_{\cal O} = 2 \;\; {\rm for} \;\; I=J=0 ,& & \;\;\; C_{\cal O} = 1/2
\;\; {\rm for} \;\; I=J=1 .
\label{eq:ratios}
\end{eqnarray}

It is easy to understand Eq. (\ref{eq:ratios})
from the point of view of the hedgehog
models themselves.  Consider a generic operator constructed from two pion
fields including any number of derivatives.  Such an operator can always be
written as \mbox{${\cal O}=h_{ab} X_a Y_b +h.c.$}, where $X$ and $Y$
are operators composed of one pion
field and any number of spatial derivatives and $a$ and $b$ are isospin
indices.
To leading order in $N_c$ , $X$ and $Y$ in the hedgehog model are given by
\cite{ANW83,CB86}:
\begin{equation}
X_a = c_{ai} X_i^{hh}, \;\; Y_a = c_{ai} Y_i^{hh},
\label{eq:collective}
\end{equation}
where $X^{hh}$ and $Y^{hh}$ are the mean-field hedgehog expressions for
the fields, and $c_{ai}$ is
the collective isorotation operator described in Refs.
\cite{ANW83,CB86}.  We wish to consider
nucleon matrix elements, thus ${\cal O}$ must be either isovector or
isoscalar and $h_{ab}$
is either $\epsilon_{abc}$ or $\delta_{ab}$. Using the properties of
the collective matrix elements and
properties of the $SU(2)$ group it is straightforward to demonstrate that

\begin{equation}
\langle N\mid {\cal O}^{I=0}\mid N \rangle = \langle N\mid X_i Y_i +
h.c. \mid N \rangle = 2 X^{hh}_{i} Y^{hh}_{i}
\label{eq:coll0}
\end{equation}
and spatial integrals of this operator will only be nonvanishing if
${\cal O}^{I=0}$ is
scalar. One can explicitly evaluate the $XY$
products in terms of collective intermediate states, and isolate contributions
from intermediate collective $N$ and $\Delta$ states:

\begin{eqnarray}
\sum^{}_{N'} \langle N\mid X_{a}\mid N^\prime \rangle \langle N^\prime
\mid Y_{a}\mid N \rangle + h.c. &=& 2/3 \;X^{hh}_{i} Y^{hh}_{i} , \nonumber \\
\sum^{}_{\Delta } \langle N\mid X_{a}\mid \Delta \rangle \langle \Delta
\mid Y_{a}\mid N \rangle + h.c. &=&  4/3 \;X^{hh}_{i} Y^{hh}_{i} .
\label{eq:coll1}
\end{eqnarray}
Thus, in agreement with Eq. (\ref{eq:ratios}), the nucleon intermediate
state accounts for
1/3 of the total in this scalar-isoscalar channel and the $\Delta$ for
2/3 of the total.
Similarly, one finds that for isovector operators
the nucleon intermediate states give 2/3 of the total, and the $\Delta$ for
1/3.

Next, let us consider how large are the $\Delta$ contributions
for physical values of $m_{\pi}$ and $N$-$\Delta$ mass splitting.
Let us introduce
\begin{equation}
d = \frac{M_{\Delta }-M_{N}}{m_{\pi}} .
\label{eq:d}
\end{equation}
$\chi PT$ implicitly assumes $d \rightarrow \infty$, while
the large-$N_c$ limit used in hedgehog models assumes
$d \rightarrow 0$. In nature,
neither of these extremes is true: $d \simeq 2.1$, which
raises troubling questions about the validity of both approaches.  It seems
plausible that a more useful way to organize the problem is to include both
the nucleon and $\Delta$ explicitly and to expand by assuming that both
$m_{\pi }$ and
$M_{\Delta }-M_{N}$ are much smaller than other scales in the problem
but with no prejudice
as to their relative size. This is in fact the spirit of the works of Jenkins
and
Manohar \cite{JM}. We note that leading singularities in this new expansion
should also come from one-pion-loop diagrams with both nucleons or
deltas are included
in the intermediate states.  Moreover, to determine the contribution to the
leading
nonanalytic behavior it is legitimate to ignore the recoil of the
baryon and use nonrelativistic baryon
propagators.

It is useful to compare the contribution to some quantity ${\cal O}$ of
diagrams
with the $\Delta $ intermediate state, ${\cal O}_{\Delta }$, to ${\cal
O}_{N}$, the contribution with the nucleon in
the intermediate state.  One can write this in the form:

\begin{equation}
\frac{{\cal O}_{\Delta }}{{\cal O}_{N}} =
\left ( \frac{g_{\pi N \Delta}}{g_{\pi NN}}
      \frac{g^{hh}_{\pi NN}}{g^{hh}_{\pi N \Delta}} \right )^{2}
 C_{{\cal O}} S_{{\cal O}}(d),
\label{ratio}
\end{equation}
where the first factor is corrects for the fact that in nature the ratio of
the $\pi $ coupling to the $\Delta $ need not be what is in the
hedgehog models to leading
order in the $1/N_c$ expansion (although in practice this ratio is within a few
percent of unity), $C_{{\cal O}}$ is a factor which only depends on the
quantum numbers
of ${\cal O}$, and $S_{{\cal O}}(d)$ is a ``$\Delta $ mass suppression
factor'' which is normalized to be
unity at $d=0.$  The spin-isospin factor $C_{{\cal O}}$ is defined in
Eq. (\ref{eq:ratios}).
Somewhat surprisingly, all three quantities considered in this note, the
isovector magnetic radius, the electric polarizability and $d^{2}
M_{N}/d(m^{2}_{\pi })^{2}$
all have the same $\Delta $ mass suppression factor:
\begin{equation}
S(d) = \frac{4}{\pi} \left\{ \begin{array}{ll}
   {\rm Arctan}  \left( \sqrt{ \frac{1-d}{1+d} } \right) / \sqrt{1-d^2}
& \;\;{\rm for} \; d \leq 1 \\
   {\rm Arctanh} \left( \sqrt{ \frac{d-1}{1+d} } \right) / \sqrt{d^2-1}
& \;\;{\rm for} \; d > 1
   \end{array} \right.
\label{eq:masssupr}
\end{equation}
This function is plotted in Fig. I.  In the case of conventional $\chi PT$
we have \mbox{$S(d \rightarrow \infty)=0$},  while for the
large-$N_c$  approximation  $S(d=0)=1$.
We note that for the physical value, $d \simeq 2$ (the blob in Fig. I),
we find $S \simeq 0.5$.  This means that for scalar-isoscalar quantities (with
$C_{\cal O} = 2)$ the $\Delta$-$\pi$ intermediate states
contribute as strongly as the $N$-$\pi$ state, and
hedgehogs overestimate the total ($N+\Delta$) contribution
by a factor of $\sim 3/2$, while conventional  $\chi PT$ underestimates
it by a factor of $\sim 1/2$. Since for vector-isovector quantities
the value of $C_{\cal O}$ is four times smaller, the effect
is reduced: hedgehogs overestimate by a factor of $\sim 1.2$,
while $\chi PT$ underestimates by a factor of $\sim 0.8$.

Figure I illustrates how far we are from the chiral limit and how slowly it
is approached.  While it is formally true that as
$d \rightarrow  \infty$ , $S \rightarrow  0,$ the
falloff is very slow, $S(d) \sim log(d)/d$. Even when
$d \sim 10$, the $\Delta$-$\pi$ contribution to
scalar-isoscalar quantities is still $\sim 40\%$  of the nucleon contribution!

In summary, we have shown that the leading nonanalytic behavior
for certain observables in large-$N_c$ hedgehog models agrees with
leading order $\chi PT$ (in its conventional version) up to
an overall factor, whose origin can be traced to the role of the $\Delta$.
These $\Delta$ effects are large. Neither approach treats them properly, which
suggests the need for significant corrections in both. Our study shows how
the magnitude of these corrections can be estimated. In hedgehog models
one can make a ``quick and dirty'' fix.  One simply isolates the nonanalytic
part
of an observable, and corrects it according to Eq. (\ref{ratio}).
Also, our analysis supports the inclusion of an explicit $\Delta$ degree of
freedom
in a modified $\chi PT$, along the lines of Refs. \cite{JM}.

Support of the the National Science Foundation (Presidential Young
Investigator grant), and of the U.S. Department of Energy is gratefully
acknowledged. We thank Manoj Banerjee for many useful discussions.
One of us (WB) acknowledges a partial support of
the Polish State Committee for Scientific Research (grants 2.0204.91.01
and 2.0091.91.01).

\newpage


Table I: Comparison of hedgehog model predictions with chiral perturbation
theory for the leading nonanalitic term of selected observables.

\vspace{30pt}

\begin{table}
\begin{tabular}{cccc}
\hline
Quantity          &  I=J &  Hedgehog & $\chi PT$  \\
\hline
$\frac{d^2 M_N}{d(m^2_{\pi})^2} = \frac{d (\sigma_{\pi N} / m^2_{\pi} )
}{d(m^2_{\pi})}$&$0$&$-\frac{1}{m_{\pi}}\frac{27}{128
\pi}\frac{g^2_A}{F^2_{\pi}}$&$-\frac{1}{m_{\pi}} \frac{9}{128
\pi}\frac{g^2_A}{F^2_{\pi}}$\\
& & & \\
$\alpha_N$ & $0$ & $\frac{e^2}{4 \pi}\frac{1}{m_{\pi}} \frac{5}{32 \pi}
\frac{g^2_A}{F^2_{\pi}}$ & $\frac{e^2}{4 \pi}\frac{1}{m_{\pi}}
\frac{5}{96 \pi}\frac{g^2_A}{F^2_{\pi}}$ \\
& & & \\
$(\kappa_p - \kappa_n){\langle r^2
\rangle}^{I=1}_m$&$1$&$\frac{M_N}{m_{\pi}}\frac{3}{16
\pi}\frac{g^2_A}{F^2_{\pi}}$&$\frac{M_N}{m_{\pi}}\frac{1}{8 \pi}
\frac{g^2_A}{F^2_{\pi}}$\\
\hline
\end{tabular}
\label{tab:comp}
\end{table}

\newpage

\centerline{Figure caption}

Figure I: The $\Delta$ mass suppression factor, $S(d)$, where $d =
(M_{\Delta} - M_N) / m_{\pi}$. The blob indicates the physical point.
$S(d)$ determines the relative contribution of $\Delta$-$\pi$ to
$N$-$\pi$ states in chiral loops, up to an overall spin-isospin factor.
See Eqs. (\ref{eq:ratios},\ref{ratio}).

\end{document}